\documentclass[
    ,final            
  ]
  {aipproc}

\layoutstyle{6x9}

\begin{document}

\title{Highest Energy Cosmic Rays}

\author{Angela V. Olinto}{
  address={Department of Astronomy \& Astrophysics, Kavli Institute of Cosmological Physics\\
  The University of Chicago, 5640 S. Ellis Ave, Chicago, IL 60637, USA\\
  olinto@oddjob.uchicago.edu}
}

\begin{abstract}
We review the current state and future prospect of ultra high energy cosmic ray physics and the relationship between cosmic rays and gamma-ray astrophysics.
\end{abstract}

\maketitle


\section{Introduction}

Cosmic rays have been known to have a {\it cosmic} origin since Victor Hess took electroscopes in  balloons above 5000 m in 1912. A few decades later, Pierre Auger showed that cosmic ray primaries reach energies in excess of $10^{15}$  eV with the detection of extensive air-showers in 1938 \cite{auger38}. Since then, cosmic rays have been observed from energies of $\sim$  MeV to above $\sim 10^{20}$ eV.  Figure 1 shows a compilation of direct and indirect (via air showers)   cosmic ray observations unified into a single spectrum. The spectrum is well fit by power-laws with spectral index $\gamma \simeq 2.7$ for energies below  $\sim 10^{15}$  eV  and  $\gamma \simeq 3$ for energies above  $\sim 10^{15}$  eV,  with a varying low energy cutoff due to solar magnetic fields. The origin of cosmic rays is still a mystery soon to become a century old (see, e.g., \cite{bhatsigl,reviews} for recent reviews).

The leading theory for the origin of cosmic rays below $\sim 10^{15}$ eV is Fermi acceleration in Galactic supernova remnant shocks, but direct evidence for this model is still lacking. The best direct test of this hypothesis is likely to come from gamma-ray astronomy. If supernovae shocks are sites of cosmic ray acceleration, the emission of gamma-rays from neutral pion decay around the accelerations site should be observed. As announced by the Hess collaboration, the detection of gamma-rays from RX J1713 \cite{hess1713} and around the Galactic Center \cite{hessGC} may be the long sought after smoking guns of Galactic cosmic ray acceleration.  Detailed spectral and spatial studies of the recently discovered sites and others yet to be discovered should resolve the mystery of lower energy cosmic rays within the next decades.

Cosmic rays at the highest energies give another opportunity to resolve the mysterious origin of cosmic rays. At energies above $\sim 10^{19} eV$ magnetic bending of cosmic rays paths are likely to ease and  cosmic ray astronomy should become feasible. These highest energy messengers are likely to bring information of extra-galactic sources of cosmic rays that may or may nor correlate to the highest energy gamma-ray emitters. While cosmic background radiations limit the arrival of gamma-rays above $\sim 10^{13} eV$ from distant galaxies a similar limitation may only occurs for cosmic rays  above $\sim 10^{20} eV$.  Combined studies of gamma-rays and cosmic rays at the highest energies available to each type of messenger will reveal the workings of the extremely energetic universe.

\begin{figure}
  \includegraphics[height=.55\textheight]{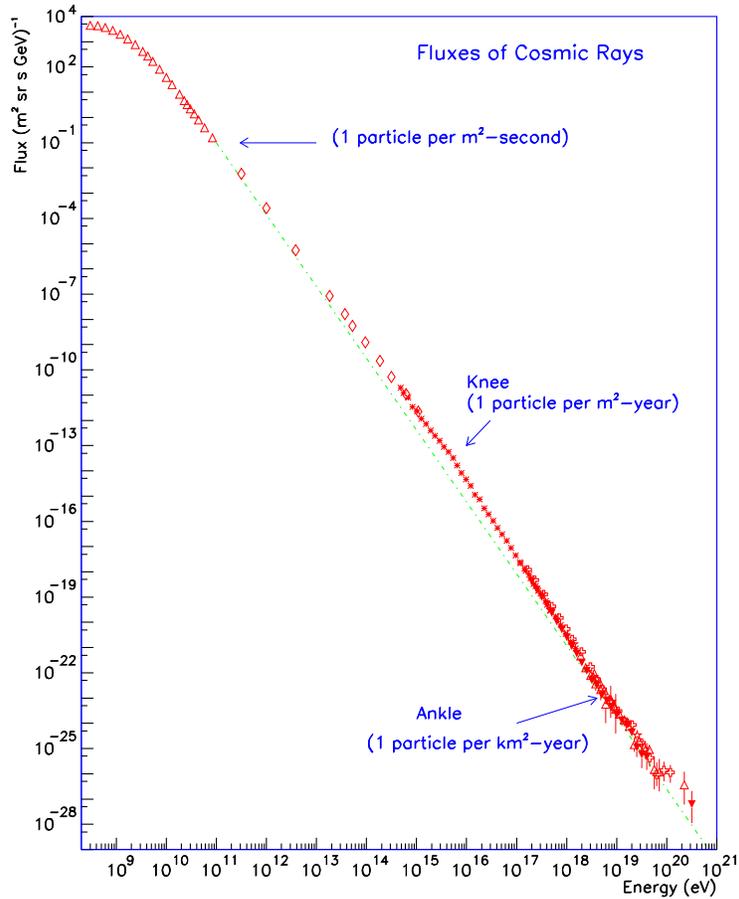}
  \caption{Spectrum of cosmic rays \cite{bhatsigl}.}
\end{figure}

\section{Cosmic Rays and Gamma-rays}

At present, the connections between gamma-ray astronomy and cosmic-rays involve the search for signatures of  cosmic ray acceleration sites, gamma-ray bounds on cosmic ray production models, and predictions of very high energy gamma-ray fluxes based on observations of ultra high energy cosmic rays (UHECRs). In addition to the search of gamma-rays from neutral pion decay around Galactic supernovae remnants, the production of  gamma-rays in extragalactic sources such as active galactic nuclei (AGN) and gamma-ray bursts may also indicate hadronic acceleration. However, spectra of AGN emission are well fit by models based on electronic processes \cite{agntalk} and the presence of hadronic acceleration is still unclear. 

Another possible region of overlap is emission from the Galactic Center \cite{melia}. Recent detections of high-energy gamma-ray emission by Whipple \cite{whipCG}, CANGAROO \cite{cangGC}, and HESS \cite{hessGC}  may indicate hadronic acceleration around the Galactic center.
These observations may help understand some initial hints of large-scale anisotropies in the cosmic ray arrival direction distribution that seem to correlate with the Galactic center region. Around  $10^{18}$ eV, the AGASA cosmic ray observatory reported a 4\% excess over the diffuse flux at the 4 $\sigma$ level \cite{agasaGC}.  HiRes \cite{hiresGC} and SUGAR \cite{sugarGC} confirm an excess with lower significance. The anisotropy points towards the Galactic center region, which may indicate a Galactic UHECR component. This small anisotropy may be associated with cosmic ray neutrons that originated either in the Galactic center region or in the Cygnus spiral arm. However, the AGASA data shows the small asymmetry  only in a narrow energy range using observations made from the northern hemisphere where the Galactic Center is obscured \cite{agasaGC}.  The construction of the Auger Observatory in the southern hemisphere \cite{auger} significantly enhances the prospects of resolving the spectral and angular shape of this asymmetry in cosmic rays.

\begin{figure}
  \includegraphics[height=.4\textheight]{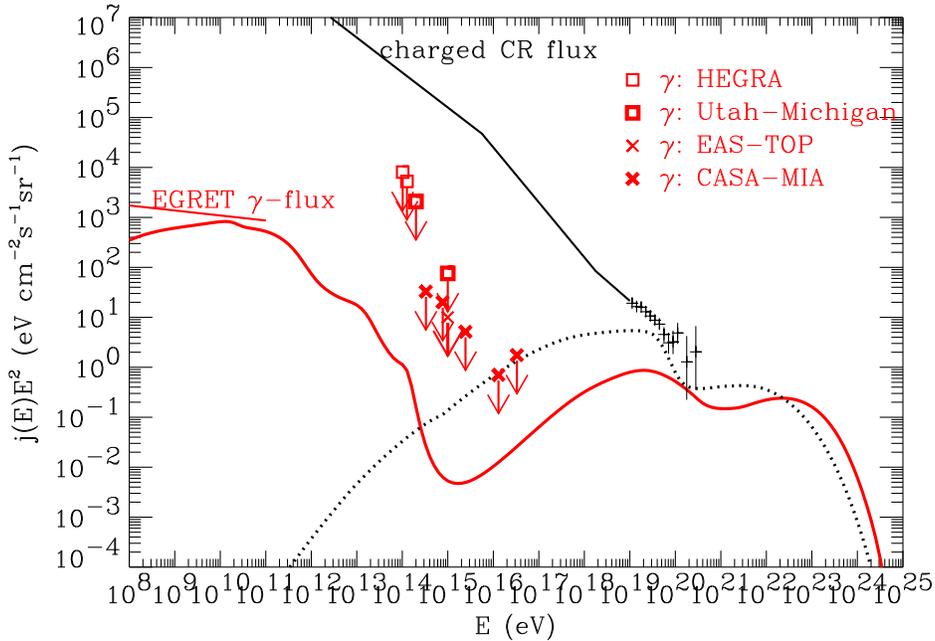}
  \caption{Predictions for the differential fluxes of
$\gamma-$rays (solid line) and protons and neutrons (dotted
line) by a topological defect model with mass scale $10^{16}\,$GeV  \cite{bhatsigl}. Also shown is the cosmic ray spectrum, gamma-ray upper limits from different experiments, and the EGRET measurement
of the diffuse $\gamma-$ray flux between 30 MeV and 100 GeV~\cite{cdkf}.}
\end{figure}

Ultra high energy cosmic ray studies have profited from lower energy gamma-ray measurements such as EGRET bounds around 10 GeV \cite{cdkf}. Explanations of the origin of ultra high energy cosmic rays based on the decay of relic objects from the early universe are usually constrained and often ruled out by the associated gamma-ray emission at EGRET energies (see, e.g., \cite{bhatsigl}). These models designed to produce UHECRs often generate a large flux of ultra high energy gamma-rays that after propagation through the universe become lower energy gamma-rays accumulating in the EGRET energy range. Figure 2 shows an example of such a UHECR model based on the decay of  topological defects  left over from a phase-transition at $10^{25} $ eV.

If sources of UHECRs are identified in the future, it is likely that these extremely high energy sources also emit high energy gamma-rays. In addition to the source emission, gamma-rays will be produced by the propagation of UHECRs through photo-pion production followed by the decay of neutral pions. 
Assuming astrophysical sources that fit the UHECR spectrum \cite{DBO03} and small scale anisotropies as indicated by AGASA \cite{agasaclusters},  a source luminosity can be derived \cite{BD04}. With these assumptions, the flux of the gamma-rays in the GeV-TeV energy range may be detected by atmospheric Cherenkov telescopes (ACTs) such as Hess and Veritas, if the intergalactic magnetic field is not larger than $\sim 10^{-10}$ Gauss \cite{FBD04}.

\begin{figure}
  \includegraphics[height=.5\textheight,angle=-90]{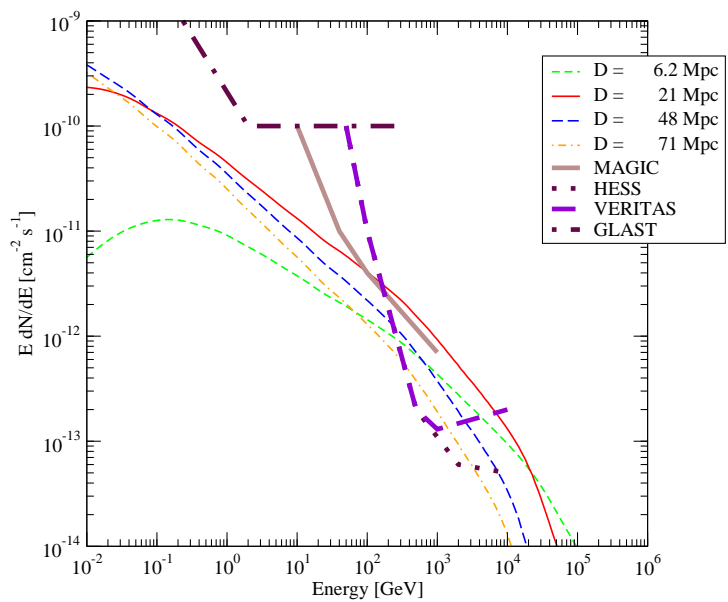}
  \caption{Flux of gamma rays at the Earth for a magnetic field
in the intergalactic medium $B=10^{-11}$~Gauss 
and a source with luminosity $2\times 10^{43}~\rm erg \, s^{-1}$ 
at energies larger than $10^{19}$ eV  \cite{FBD04}.}
\end{figure}

As UHECR observatories capable of observing large number of events above $\sim 10^{20}$ eV are being built, such as Auger and EUSO, the possibility of identifying the sources of UHECRs before the 100th anniversary of the discovery cosmic rays in 2012 is becoming a reality. These new observatories  are likely to start the new era of particle astronomy which will likely increase the connections between cosmic ray and gamma-ray astronomy. 

\subsection{Ultra High Energy Cosmic Rays}

Ultra-high energy cosmic rays are the highest energy messengers of the present universe. The highest energy cosmic photons observed thus far reach up to $10^{13}$ eV. Extragalactic photons of higher energies loose a significant fraction of their energies due to pair production in the cosmic background radiation as they traverse large regions of intergalactic space.  In contrast, cosmic rays are observed with energies as high as $3 \times 10^{20}$ eV and with fluxes well above upper limits on high-energy gamma-ray fluxes. 

However, the origin of cosmic rays remains a mystery hidden by the fact that these relativistic particles do not point back to their sources. These charged particles are deflected by magnetic fields that permeate interstellar and intergalactic space. Galactic magnetic fields are known to be around a few micro Gauss in the Galactic disk and are expected to decay exponentially away from the disk \cite{kronberg}. Intergalactic fields are observed in dense clusters of galaxies, but it is not clear if there are intergalactic magnetic fields in the Local Group or the Local Supergalactic Plane. On larger scales, magnetic fields are known to be weaker than $\sim$ 10 nano Gauss \cite{BBO}. 

\begin{figure}
  \includegraphics[height=.4\textheight]{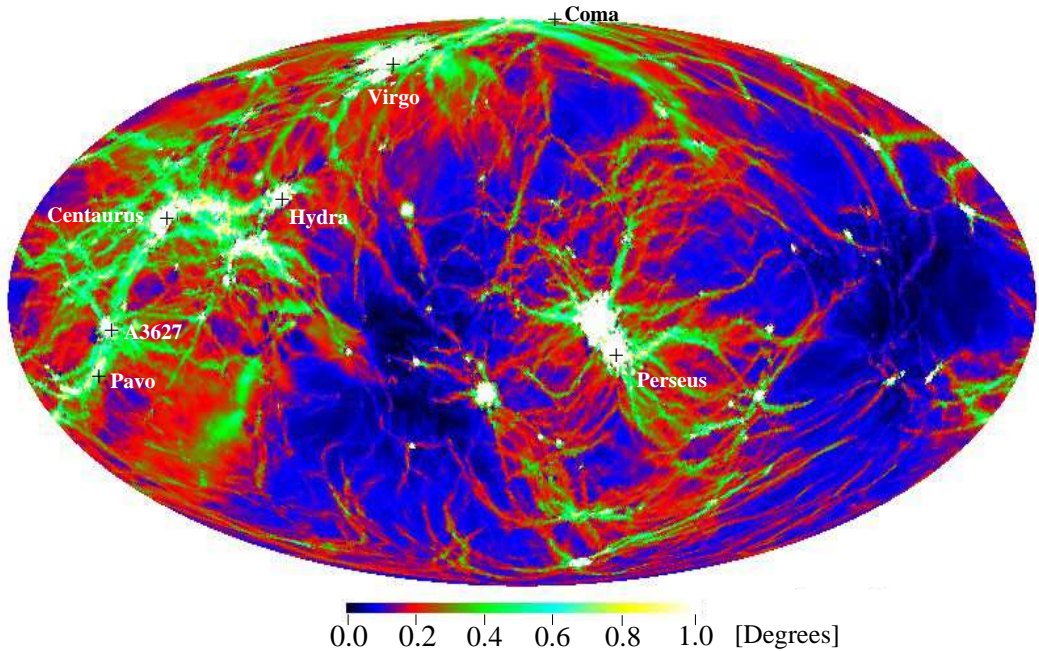}
  \caption{Shown is a full sky map of expected angular deflections for extragalactic cosmic ray sources using simulations of large scale structure formation with magnetic fields \cite{dolag}. Cosmic ray protons with arrival energy $E = 4\times 10^{19}$~eV were propagated through the whole simulation
volume within a radius of $110$~Mpc around the position of the Galaxy. The coordinate system is galactic, with the galactic anti-center in the middle of the map.}
\label{defl19.5}
\end{figure}

As cosmic ray energies reach $10^{20}$ eV per charged nucleon, Galactic and intergalactic magnetic fields cannot bend particle orbits significantly and pointing to cosmic ray sources becomes feasible. Recent high-resolution simulations of large-scale structure formation in a $\Lambda$CDM universe can follow the magnetic field evolution from seed fields to present fields in galaxies and clusters \cite{dolag}. The intergalactic medium fields in these simulations are consistent with Faraday rotation measurements at the at the $10^{-9}-10^{-8}$ Gauss level. In addition to simulating the field evolution, cosmic ray protons are propagated through a volume of 110 Mpc radius. Figure \ref{defl19.5} shows that  the deflection from the source position to the arrival direction for protons with arrival energy of  $ 4 \times 10^{19}$ eV can reach around 1 degree in the densest regions \cite{dolag}. For protons arriving with  $10^{20}$ eV the deflections are less than a$\sim$  0.1 of a degree (which is significantly smaller than the resolution of UHECR observatories) \cite{dolag}. Therefore, at ultra high energies there is finally the opportunity to begin cosmic ray astronomy.

\begin{figure}
  \includegraphics[height=.4\textheight]{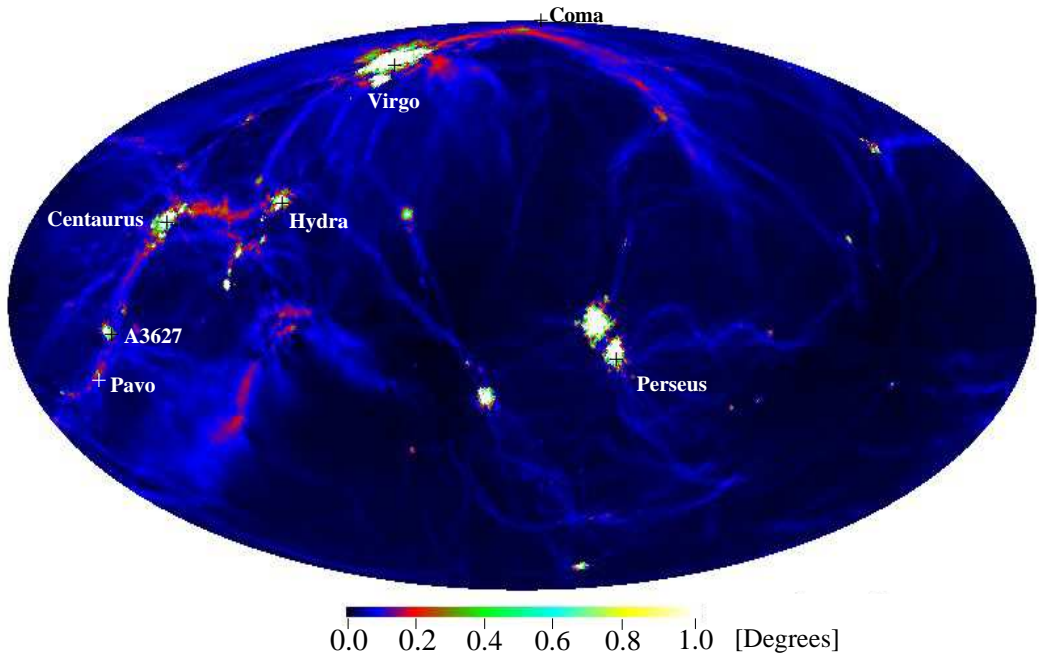}
  \caption{Shown is a map as in Fig. \ref{defl19.5} of the expected angular deflections for protons with arrival energy $E = 1\times 10^{20}$~eV including energy losses in the propagation \cite{dolag}.}
\label{defl20}
\end{figure}

In addition to the ability to point back to the source position, cosmic ray protons of energies around $10^{20}$ eV should display a well-known spectral feature called the GZK cutoff \cite{gzk}. This cutoff was proposed in 1966 by Greisen, Zatsepin and Kuzmin as a natural end to the cosmic ray spectrum due to photopion production off the then recently discovered cosmic microwave background radiation.  The presence of microwave photons through cosmic space induces the formation and subsequent decay of the $\Delta^+$ resonance for protons with energies above $\sim 10^{20}$ eV that traverse distances longer than $\sim$ 50 Mpc. The effect of photopion production is to decrease the energy of protons from distant sources resulting in a hardening of the spectrum between $10^{19}$ eV and $10^{20}$ eV followed by a sharp softening past $10^{20}$ eV. Depending on the maximum energy of ultra high-energy cosmic ray sources and their distribution in the universe, the spectrum may harden again past the GZK feature displaying the injected spectrum of nearby sources.

The search for the origin of the highest energy particles is being undertaken by a number of experiments. At present, observations of cosmic rays at the highest energies have yielded measurements of the spectrum, arrival direction distribution, and composition of UHECRs below $10^{20}$ eV.   The cosmic ray spectrum past $10^{20}$ eV should show the presence or absence of the GZK feature, which can be related to the type of primary (e.g., protons) and source (injection spectrum and spatial distribution) of UHECRs. Currently, the two largest exposure experiments, the Akeno Giant Airshower Array (AGASA) and the High Resolution Fly's Eye (HiRes) have conflicting results at the highest energies (above $\sim 10^{20}$ eV ) where limited statistics and systematic errors prevent a clear resolution. 

\begin{figure}
  \includegraphics[height=.5\textheight]{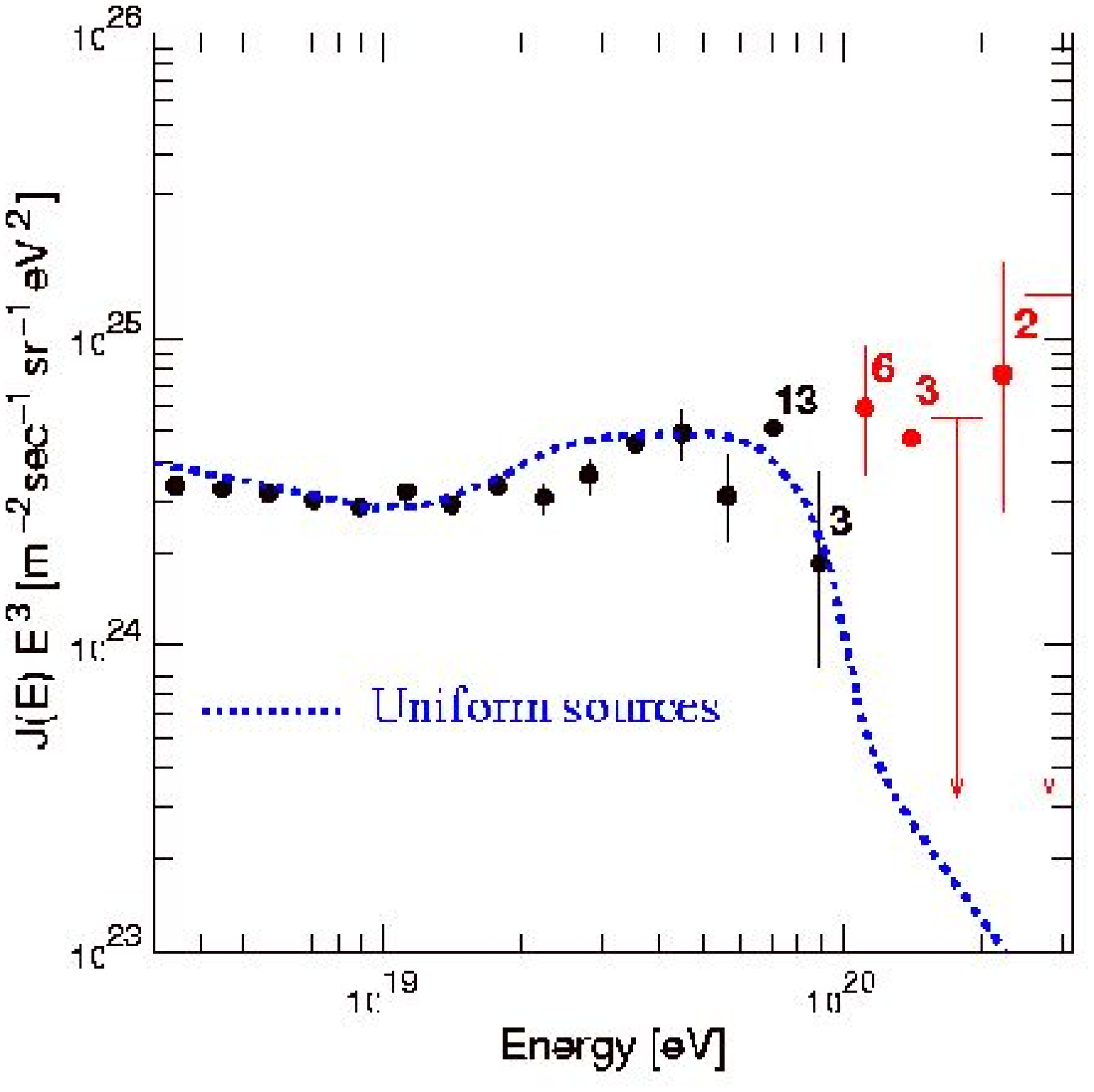}
  \caption{AGASA spectra \cite{agasaspec}.}
  \label{agasaspec}
\end{figure}

AGASA is a 100 km$^2$ ground array of scintillator and muon detectors. AGASA data shows a distribution of arrival directions which is mainly isotropic with an indication of clustering of cosmic rays at the highest energies and smallest angles \cite{agasaclusters}. In addition, the spectrum shows the lack of a GZK cutoff around $10^{20} $ eV (see figure \ref{agasaspec}).  The flux above $10^{20}$ eV does not show the expected  GZK cutoff with the detection of  11 Super-GZK events, i.e., 11 events with energies above $10^{20}$ eV \cite{agasaspec}. These findings argue against the notion of extragalalactic proton sources of UHECRs and for a  unexpected new source at the highest energies.

In contrast, the HiRes monocular spectrum indicates smaller fluxes past $10^{20}$ eV which is consistent with a GZK feature \cite{hiresspec}. HiRes reports only two events with energies above $10^{20}$ eV (1 seen in Stereo). HiRes is composed of fluorescence telescopes built in two different sites in the Utah desert to be used as a stereo fluorescence detector. While stereo results do not have comparable exposure to AGASA yet, monocular data do have comparable exposure. Mono HiRes analysis shows no evidence of clustering of arrival directions on small scales and a decrease in flux consistent with the GZK feature. In addition to the spectrum and distribution of arrival directions, HiRes data indicates that between $10^{18}$ eV and $10^{19.3}$ eV the composition shifts from a heavier (iron dominated) component to lighter (proton dominated) component.

The implications of the differing results from AGASA and HiRes are especially intriguing at the highest energies. The  discrepancies between HiRes and AGASA spectra corresponds to  $\sim$ 30\% systematic error in energy scales.  Possible sources of systematic errors in the energy measurement of the AGASA experiment were comprehensively studied to be at around 18 \% \cite{agasasys}.  Systematic errors in HiRes are still being evaluated, but are likely to be dominated by uncertainties in the absolute fluorescence yield, the atmospheric corrections, and the calibration of the full detector., which could amount to at least $\sim$ 20\% systematic errors in energy calibration.

\begin{figure}
  \includegraphics[height=.4\textheight]{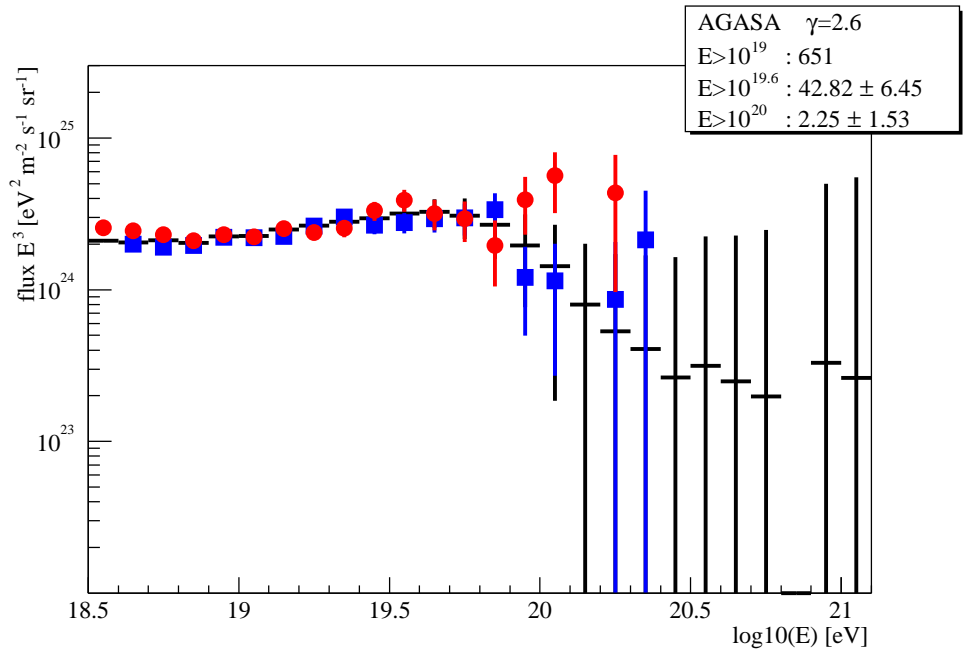}
  \caption{AGASA with -15\% energy shift and HiRes with +15\% shift \cite{DBO03}.}
  \label{statsys}
\end{figure}

Although control of systematic errors is crucial, the statistics accumulated by both HiRes and AGASA are not large enough for a clear measurement of the GZK feature. Figs. \ref{statsys} shows the range of 400 simulated spectra of protons propagating in intergalactic space with injection spectral index of 2.8 for AGASA  and 2.6 for HiRes exposures. In addition, the data from AGASA and HiRes are shown with a systematic energy shift of -15\% for AGASA and +15\% for HiRes \cite{DBO03}.  The disagreement between the two experiments is only  about 2 $\sigma$ using these arbitrarily chosen  systematic corrections, which are well within the possible range of systematic errors.  

\begin{figure}
  \includegraphics[height=.5\textheight]{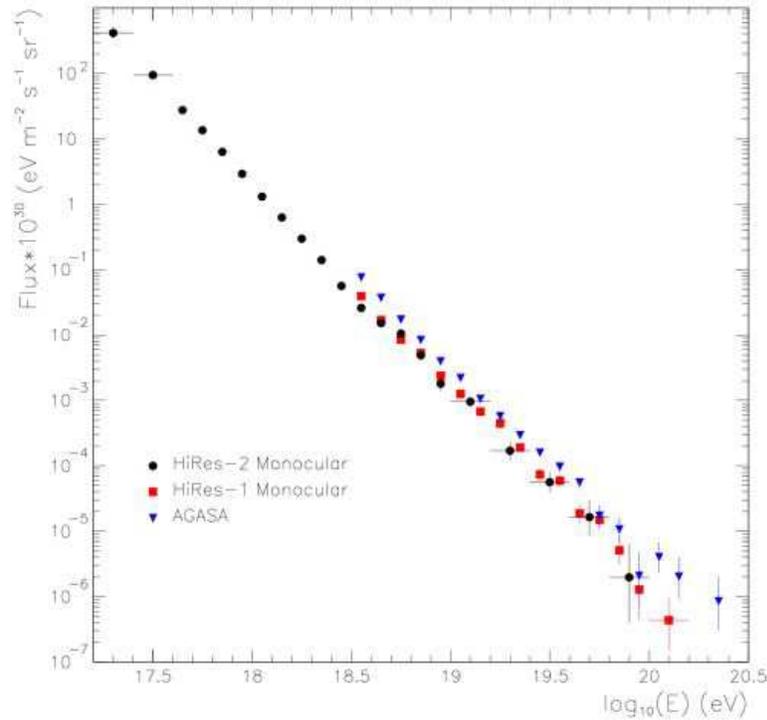}
  \caption{AGASA and HiRes spectra \cite{Bergman}.}
  \label{agasahires}
\end{figure}

The systematic energy shifts between AGASA and HiRes through the range of observed energies is more easily seen when the two spectra are plotted on a flux versus energy plot (see Figure \ref{agasahires}). In addition, the discrepancies between the two experiments are not as accentuated as in the traditional plots of flux times $E^3$. Finally, the low exposure above  $10^{20}$ eV of both experiments prevents an accurate determination of the GZK feature or lack of it.  The lessons for the future are clear: improve the statistics significantly above  $10^{20}$ eV and understand the sources of systematic errors.

\section{Preview of the Next Generation}

Neither AGASA nor HiRes have the necessary statistics and control of systematics to determine in a definitive way the existence of either the GZK feature or of a novel source of Super-GZK events. Moreover, if the AGASA clusters are an indication of point sources of UHECRs, a large number of events per source will be necessary to study their nature. In order to discover the origin of UHECRs, much larger aperture observatories are now under construction, i.e.,  the Pierre Auger Project\cite{auger}, or under development, i.e., the Telescope Array\cite{TA} and the Extreme Universe Space Observatory\cite{EUSO}. 

The Pierre Auger Project  will consist of two giant airshower arrays one in the South and one in the North each with 1600 water Cherenkov detectors covering 3000 km$^2$ and four sites of fluorescence telescopes.  Auger is being built to determine the spectrum, arrival direction, and composition of UHECR in a full sky survey. The survey should provide large event statistics and control of systematics through detailed detector calibration of the surface array and fluorescence detectors individually in addition to the cross-calibration of the two detection techniques through the observation of hybrid and stereo-hybrid events. Depending on the UHECR spectrum, Auger should  measure the energy, direction and composition of about 60 events per year above 10$^{20}$ eV and about 6000 events per year above 10$^{19}$ eV.  In addition, it should be able to detect a few neutrino events per year if the UHECRs are extragalactic protons.

\begin{figure}
  \includegraphics[height=.35\textheight]{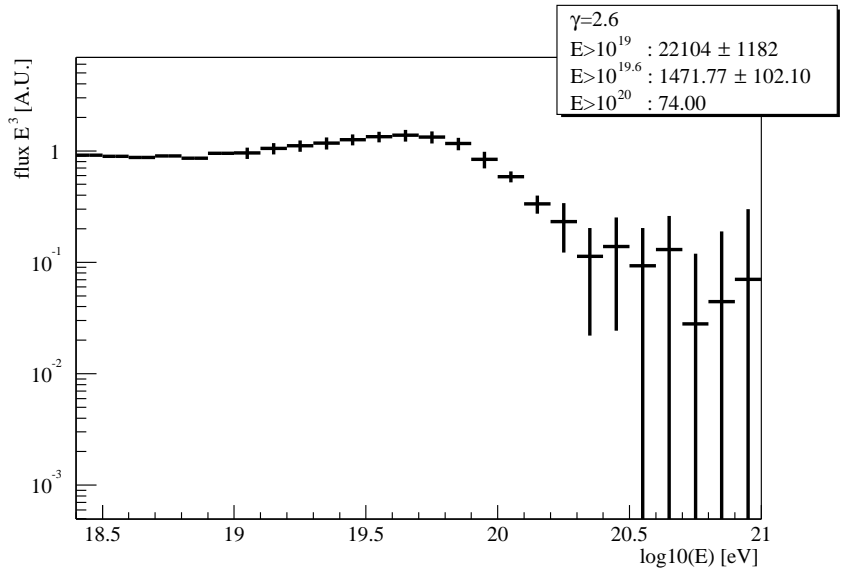}
  \caption{Auger South statistics at the GZK feature [26].}
\end{figure}

The Auger surface array  is composed of stand alone 1.5 meter tall water tanks that are powered by solar cells, timed by GPS systems, and  communicate via radio antennas. Three photomultipliers per tank register the Cherenkov light when shower particles cross the tanks. Having three photomultipliers per tank allows the self-calibration of each tank in the field.  The height of the tanks makes the ground array an excellent detector for inclined showers. Inclined showers and their asymmetries allow for a novel method for composition studies and for the detection of neutrino showers from horizontal and Earth skimming high energy neutrinos.

The fluorescence detectors at the Auger observatory have a complete calibration system. The atmospheric monitoring includes lasers, lidars, ballon radio sondes, cloud monitors, and movable calibration  light sources [44,45]. In addition, the whole telescopes including mirrors are calibrated from front to end with light sources. 

Hybrid detection is a powerful measurement of individual showers and can be used to reach large statistics on energies down to 10$^{18}$ eV with the use of fluorescence and a small number of tanks per event [42].  The ability to study events at 10$^{18}$ eV  in the Southern hemisphere will be crucial in confirming the reported anisotropies toward the Galactic Center region.  The combination of mono fluorescence events that triggered even a single tank allows for great angular reconstruction of events comparable to stereo events  [43].

The Auger collaboration consists of about 250 scientists from 16 countries. The Southern Auger Observatory is already operating with 500 surface detector tanks deployed and two fluorescence telescope sites completed. The first science results of the observatory should be presented in the Summer of 2005.

Another upcoming experiment is the recently approved Telescope Array (TA) which consists of  a hybrid detector of three fluorescence telescopes overlooking a scintillator array. The array would cover about 400 km$^2$ with 1.2 km spacing. The design limits the exposure at theh ighest energies but is suited to energies from $\sim 10^{17}$ eV to  $\sim 10^{20}$ eV, where a transition between Galactic and extragalactic UHECRs are expected.  TA should be able to see some super-GZK events but with significantly smaller statistics than the Auger project. Instead, TA is planning to concentrate their efforts in having a broad reach in energies to study the spectrum and composition through the transition from Galactic to extragalactic that may involve a simultaneous heavy to light primaries transition.

\begin{figure}
  \includegraphics[height=.3\textheight]{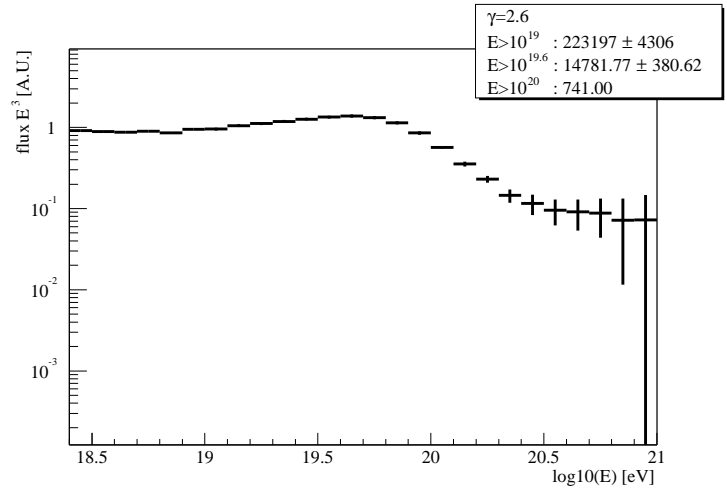}
  \caption{EUSO statistics at the GZK feature \cite{DBO03}.}
  \label{eusospec}
\end{figure}

Finally, the Extreme Universe Space Observatory (EUSO) is a fluorescence detector designed for the International Space Station (ISS) aiming at observations of extremely high energy cosmic-rays (EHECRs), i.e., cosmic rays between $10^{20}$ and $10^{22}$ eV. EUSO will observe showers from above the atmosphere and will have full sky coverage due to the ISS orbit. This project is a good complement to ground arrays since, it will focus on larger energy scales and will have  different systematic effects. Their threshold may be above $5 \times 10^{19}$ eV depending on technical features of the fluorescence detectors. The telescope's expected angular resolution is  $\sim$ 0.2 degrees and the energy resolution  about $\sim$ 20\%. The aperture may reach $3 \times 10^6$  km$^2$-sterad with a 10\% duty cycle. This can translate into about 3000  events per year for energies above $10^{20}$  eV (see Fig. \ref{eusospec}). 

\section{Conclusion}

After decades of attempts to discover the origin of ultra-high energy cosmic rays, present results are still inconclusive. The results from past experiments show the need to understand and control systematic effects within each technique and to cross-calibrate the two techniques presently available for UHECR studies (ground arrays and fluorescence). In addition, the lack of sufficient statistics limits the discussion of an excess flux or a drop in flux around the GZK feature. Next generation experiments are gearing up to accumulate the necessary statistics while having a better handle on the systematics. In the following decade, we may see the growth of a new astronomy with ultra-high energy charged particles and finally resolve the almost century old puzzle of the origin of cosmic rays.

\begin{theacknowledgments}
This work was supported in part by the KICP under NSF PHY-0114422 , 
by  the NSF through grant AST-0071235, and the DOE grant DE-FG0291-ER40606. 
\end{theacknowledgments}

\end{document}